\documentclass[preprint]{aastex}
\usepackage{graphicx}
\usepackage{epstopdf}
\usepackage{natbib}
\begin{document}

\title{NEOWISE observations of comet C/2013 A1 (Siding Spring) as it approaches Mars}

\author{R. Stevenson\altaffilmark{1}, J. M. Bauer\altaffilmark{1,2}, R. M. Cutri\altaffilmark{2}, A . K. Mainzer\altaffilmark{1}, F. J. Masci\altaffilmark{2}}

\altaffiltext{1}{Jet Propulsion Laboratory, California Institute of Technology, 4800 Oak Grove Drive, Pasadena, CA 91109}
\altaffiltext{2}{Infrared Processing and Analysis Center, California Institute of Technology, Pasadena, CA 91125}

\begin{abstract}

The Near-Earth Object Wide-field Infrared Survey Explorer (NEOWISE) mission observed comet C/2013 A1 (Siding Spring) three times at 3.4~$\mu$m and 4.6~$\mu$m as the comet approached Mars in 2014. The comet is an extremely interesting target since its close approach to Mars in late 2014 will be observed by various spacecraft in-situ. The observations were taken in 2014 Jan., Jul. and Sep.\ when the comet was at heliocentric distances of 3.82~AU, 1.88~AU, and 1.48~AU. The level of activity increased significantly between the Jan.\ and Jul.\ visits but then decreased by the time of the observations in Sep., approximately 4 weeks prior to its close approach to Mars. In this work we calculate Af$\rho$ values, and CO/CO$_{2}$ production rates.

\end{abstract}

\keywords{comets:individual; C/2013 A1 (Siding Spring)}

\section{Introduction}

Comet C/2013 A1 (Siding Spring) presents both risk and opportunity at its close approach to Mars on 2014 Oct.\ 19. It will pass approximately 135,000 km $\pm$ 5000 km from the planet's center \citep{2014ApJ...790..114F} - close enough for detailed observations by spacecraft at Mars but also close enough for ejected dust and gas to reach Mars' atmosphere, potentially affecting spacecraft in orbit. The potential risk posed by the dust has been downgraded through careful dynamical modeling. Early studies suggested relatively high amounts of dust would reach the Martian atmosphere \citep{2014ApJ...787..115Y,2014Icar..231...13M}. Later studies that included lower ejection velocities and  radiation pressure effects showed 
%\citep{2014ApJ...787L..35T,,2014ApJ...792L..16K} that showed
 only a low fluence of old, larger dust grains is expected to reach the atmosphere \citep{2014ApJ...787L..35T,2014ApJ...792L..16K}. However, it is possible that the gas coma may sufficiently excite the atmosphere of Mars to cause increased drag on orbiting satellites \citep{2014Icar..237..202Y}. The comet is also interesting in its own right as a long period comet on a near-parabolic retrograde orbit, bringing primitive material in from the Oort Cloud.

In this paper we use data from the Near-Earth Object Wide-Field Infrared Survey Explorer reactivation (NEOWISE; \citealt{2014ApJ...792...30M}) mission to characterize the comet's activity on three occasions in 2014 Jan.\, Jul.\, and Sep.\ as the comet approached Mars. We derive dust and gas production rates using near-infrared wide-field images and examine the evolving morphology of the coma.

\section{Data Acquisition and Reduction}

The NEOWISE mission utilizes the Wide-Field Infrared Survey Explorer (WISE) spacecraft \citep{2010AJ....140.1868W}, which completed an all-sky survey at four wavelengths of 3.4, 4.6, 12, and 22~$\mu$m from January to August 2010 using a 40~cm cryogenically-cooled telescope. After the depletion of the spacecraft's solid hydrogen, the mission could still operate at the two shortest wavelengths of 3.6 and 4.6~$\mu$m. This phase of the mission was renamed NEOWISE \citep{2011ApJ...731...53M} and continued for several months until Feb 1, 2011 after which the spacecraft was placed into hibernation. The spacecraft was then reactivated in Dec 2013 for the start of the NEOWISE reactivation mission \citep{2014ApJ...792...30M}. The telescope is conducting an all-sky survey in a low-Earth orbit, taking images of the sky every 11 s at a solar elongation of $\sim$92$^{\circ}$.5. The images have an effective exposure time of 7.7 s, and effective point-source full-width half-maxima (FWHM) of 6$^{\prime\prime}$.25. There is approximately 10\% overlap between successive frames, allowing for the linking of moving object detections.   The images, which are 47$^{\prime}$ on a side, undergo instrumental and astrometric calibration, followed by source extraction at at the Infrared Processing and Analysis Center (IPAC) at the California Institute of Technology  \citep{2010AJ....140.1868W}. All data from the WISE prime mission in 2010 have been publicly released via NASA's Infrared Science Archive (IRSA)\footnote{http://irsa.ipac.caltech.edu/}, and the first data release from the NEOWISE reactivation mission will be in March 2015 through IRSA.  

%Typical astrometric residuals are 0$^{\prime\prime}$.67 and XX, respectively.

%NEOWISE has restarted and provided excellent opportunity to monitor the comet as it approaches Mars. Three sets of observations. Restart paper - Mainzer et al. \citep{2014ApJ...792...30M}. Restart of the WISE mission at two shortest wavelengths of 3.4 and 4.6 microns. Field of view is 47' on a side. Spatial resolution of FWHM of 6.25 arc seconds. Scans at 92.5 deg solar elongation, which allows it to have multiple encounters/visits/epochs of observations for the same Solar System object, if it is moving in the direction of the scanning telescope. With 10$\%$ overlap between frames. Scan path progresses 1 deg per day. Will do this over the 3 year lifetime of its mission. 

%``Scans continuously along great circles with approximately constant ecliptic longitude, while a scan mirror freezes the sky on the focal planes''. Exposure time is 7.7 s. Data processing and archiving at IPAC.

% "before being passed through WMOPS. WMOPS uses a reference sky to reject stationary sources and link detections into tracklets." Is this necessary? Need to say how comet was identified.

The motion of comet Siding Spring meant that it was observed by NEOWISE on three separate occasions, hereafter referred to as ``visits'',  on 2014 Jan.\ 16-17, Jul.\ 28-29, and Sep.\ 21-22. Observing details are given in Table~\ref{table:obs}.
% The heliocentric distance of the comet was 3.82, 1.88, and 1.48 AU in the Jan.\, Jul.\, and Sep.\ visits, respectively. 

%On each visit, the comet was initially identified by the WISE Moving Object Processing System (WMOPS) and 

\begin{deluxetable}{cccc}
\tabletypesize{\scriptsize}
\tablecaption{Observations}
\tablewidth{0pt}
\tablehead{
\colhead{Quantity} & \colhead{January Visit} & \colhead{July Visit} & \colhead{September Visit}}
\startdata
%Date [UT] & 2014-01-16 02:09 - 2014-01-17 01:52 & 2014-07-28 08:10 - 2014-07-29 01:32 & 2014-09-XX XX:XX - 2014-09-XX XX:XX \\
Date (UT)\tablenotemark{a} & 2014 Jan.\ 16.62 & 2014 Jul.\ 28.87 & 2014 Sep.\ 21.73 \\
%56673.0893885 - 56674.0776535 & 56866.3399642 - 56867.0637513 & 56866.3399642 - 56867.0637513\\
N\tablenotemark{b} & 16 & 11 & 9 \\
$\Delta$\tablenotemark{c} (AU) & 3.67 & 1.54 & 1.04 \\
r$_{H}$\tablenotemark{d} (AU) & 3.82 & 1.88 & 1.48 \\
$\alpha$\tablenotemark{e} ($^{\circ}$) & 14.9 & 32.7 & 42.6 \\
Image scale (km arcsec$^{-1}$) & 2662 & 1117 & 754 \\
3.4~$\mu$m flux (mJy) & 0.5 $\pm$ 0.1 & 7.4 $\pm$ 1.6 & 18 $\pm$ 4 \\
4.6~$\mu$m flux (mJy)  & 0.6 $\pm$ 0.1 & 21 $\pm$ 5 & 74 $\pm$ 16 \\
Af$\rho$ (cm) & 432 $\pm$ 21 & 726 $\pm$ 40 & 724 $\pm$ 40 \\
CO (molecules s$^{-1}$) & (3.56 $\pm$ 0.36) $\times$ 10$^{27}$ & (1.65 $\pm$ 0.17) $\times$ 10$^{28}$ & (1.25 $\pm$ 0.13) $\times$ 10$^{28}$ \\
CO$_{2}$ (molecules s$^{-1}$) & (3.42 $\pm$ 0.34) $\times$ 10$^{26}$ & (1.56 $\pm$ 0.16) $\times$ 10$^{27}$ & (1.18 $\pm$ 0.12) $\times$ 10$^{27}$ \\
\enddata
\tablenotetext{a}{Mid-point of median stack}
\tablenotetext{b}{Number of individual images used}
\tablenotetext{c}{Geocentric distance at mid-point}
\tablenotetext{d}{Heliocentric distance at mid-point}
\tablenotetext{e}{Phase angle at mid-point}
\label{table:obs}
\end{deluxetable}

\begin{figure}[h]
\centering
%\subfloat{\includegraphics[totalheight=4.24cm]{C2013A1_jan14_export.eps}}
%\hspace{2bp}
%\subfloat{\includegraphics[totalheight=4.24cm]{C2013A1_jul14_export.eps}}
%\hspace{2bp}
%\subfloat{\includegraphics[totalheight=4.24cm]{C2013A1_sep14_export.eps}}
\includegraphics[totalheight=4.24cm]{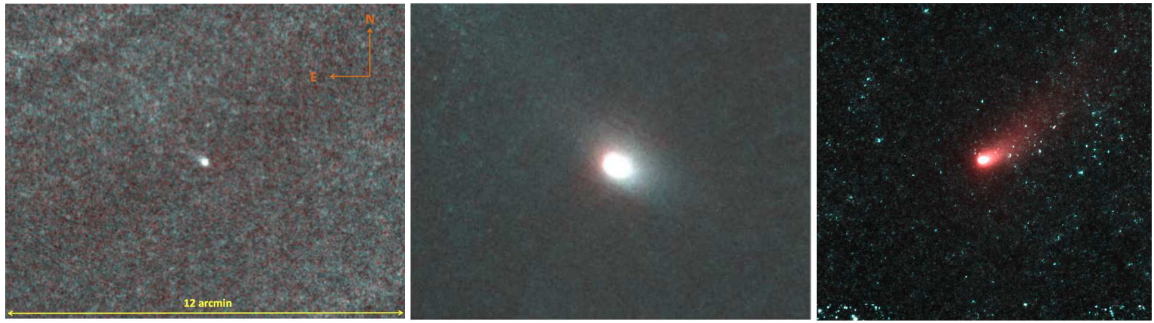}
%\caption{}
\caption{Two-band false-color images of Comet C/2013 A1 (Siding Spring) as observed by the NEOWISE mission in 2014 January (left panel), July  (middle panel), and September (right panel). The 3.4 $\mu$m band is shown as blue, while the 4.6 $\mu$m band is shown as red. The contrast and stretch of each image is adjusted for clear viewing and are not constant across the panels. However, the scale and orientations, as indicated in the left panel, are the same for all images.}
\label{fig:C2013A1images}
\end{figure}

We used predicted positions from the Jet Propulsion Laboratory's Horizons service\footnote{http://ssd.jpl.nasa.gov/horizons.cgi} and the Moving Object Search Tool\footnote{http://irsa.ipac.caltech.edu/applications/MOST/} \citep{2012wise.rept....1C} to identify the images covering comet Siding Spring. 
%Sixteen images were obtained in Jan.\, 11 in Jul.\, and 9 in Sep, with successive images typically separated by $\sim$ 1.5 hours and each visit's dataset covering between 16.7 - 23.7 hours in total.
 Trailing during each 7.7~s exposure was at most under than 1$^{\prime\prime}$, which is significantly less than the FWHM. %Astrometric measurements of comet Siding Spring were reported to the Minor Planet Center. 
We coadded the images of the comet in the moving reference frame for each visit using the ICORE image co-addition package that includes outlier-rejection \citep{2013ascl.soft02010M}. This improved the signal-to-noise ratio and resampled the pixel scale from 2$^{\prime\prime}$.75 pixel$^{-1}$ to 1$^{\prime\prime}$.0 pixel$^{-1}$. The coadded images were centered on the predicted position of the comet. Two-band color images of comet Siding Spring from each visit are shown in Figure~\ref{fig:C2013A1images}. The comet appeared active at each visit.

%astrometry reported to the Minor Planet Center.

\begin{figure}[h]
\centering
\includegraphics[totalheight=4.1cm]{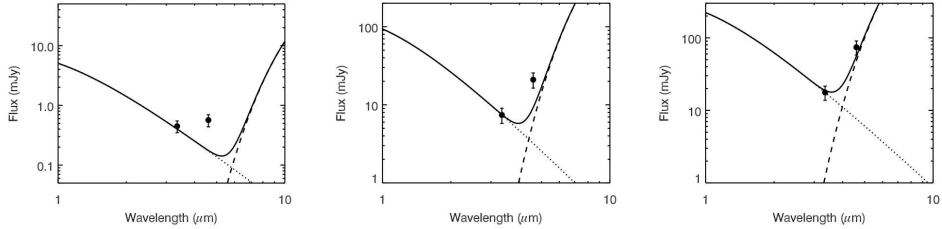}
\caption{Reflected light (dotted line) and thermal emission (dashed line) fits of Comet C/2013 A1 (Siding Spring) as observed by the NEOWISE mission in 2014 Jan.\ (left panel), Jul.\ (middle panel), and Sep.\ (right panel). The fluxes measured are shown as points and the sum of both models is shown by the solid line.}
\label{fig:C2013A1SEDs}
\end{figure}

We used circular apertures with radii of 11$^{\prime\prime}$ to perform photometry on each coadded image. Counts were converted to fluxes using instrumental zero points using the same method as used for WISE catalog source data \citep{2010AJ....140.1868W}. The background was estimated and subtracted using the modal value in an aperture of at least 1.4 square arcminutes located $\geq 1^{\prime}$ from the nucleus, and far from the coma. The dominant source of error is the uncertainty on the absolute calibration. We fit a reflected light model to the observed data at 3.4 $\mu$m, making the assumption that the flux observed is due to reflected light from dust grains, the grains have a neutral reflectance. The reflected light is thus effectively the solar flux scaled to the 3.4$\mu$m signal. We neglected light reflected or emitted by the nucleus, the size of which is unknown at this time because it has been obscured by dust since discovery. If the nucleus is on the order of 1~km in radius, it would contribute a few tenths of a $\mu$Jy to the fluxes observed during the 2014 Jan.\ visit when the comet was least active and the nucleus likely contributed the largest fraction of the light. This is significantly less than the uncertainty in the flux measurements and we conclude that the nucleus contributes a negligible amount of light. We also computed a theoretical thermal contribution to the flux by calculating the blackbody radiation that would be emitted by the quantity of dust estimated from the 3.4~$\mu$m flux. We used a Planck function and assumed an emissivity$\sim$0.9, so that the temperature of the grains scaled as $286K \times r_{H}^{-1/2}$.  It is entirely possible that the grain spectral energy distribution is more complicated, but the singular data point at 3.4$\mu$m only allows for a simplified dust model constraint. The measured fluxes for each visit are listed in Table~\ref{table:obs} and shown in Figure~\ref{fig:C2013A1SEDs} with the reflected light and predicted thermal light models. 

%Size of nucleus? Bodewits 2014 $>$ 0.34 km radius. Size of nucleus is unknown, since comet has been active since discovery. Estimate is X. Even if 1~km in radius, makes small difference. Calculated to be X in month Y, makes small difference.

%Aperture 60-80 centered on the nucleus for January. July: 1-40 arc sec annulus located (+200,-200) arc sec from nucleus. Sep: 0-80 arc sec aperture (-300, -200) from nucleus. Center of comet is (-17,5).

%Background subtraction: offset XX arcsec  from nucleus, aperture size and location from Gerbs.

%Jan
%background W1:      1.51240  +/-    0.102910
%background W2:      4.51464  +/-    0.126060

%Jul
%background W1:      1.46305     0.109236
%background W2:      4.49471     0.148208

%Sep
%background W1:      2.35617     0.273527
%background W2:      5.77042     0.241554

\section{Results}

%Flux fitting - scaled at 3.4 microns. Maybe this should be in data. Thermal based on what since there are no data points?

\subsection{Dust Production}

We converted the flux measured at 3.4~$\mu$m to the quantity $Af\rho$ - a proxy for the dust production rate discussed in \citet{1984AJ.....89..579A} - using equation~\ref{eq:afp}.

\begin{equation}
Af\rho =\frac{\left(2 \Delta r_{H}\right)^{2}}{\rho} \frac{F_{comet}}{F_{\odot}}
\label{eq:afp}
\end{equation}

where A is the Bond albedo of the dust at the phase angle of observation, $f$ is the filling factor of the dust grains within the aperture, $\rho$ is the aperture size in cm, $\Delta$ and $r_{H}$ are the geocentric and heliocentric distances in cm and AU, respectively, and $F_{comet}$ and $F_{\odot}$ are the flux from the comet and the Solar flux at 1~AU, respectively.
% We assume that the albedo of the cometary dust at 3.4~$\mu$m is 0.08, approximately twice that found to be typical of comets at optical wavelengths \citep{2011ApJ...742...40G,2014ApJ...791..121M,2004come.book..223L}.
 As before, we used 11$^{\prime\prime}$ radius apertures. We estimate the uncertainties on our measurements by examining the spread of results obtained when using apertures of radii 9, 11, and 22$^{\prime\prime}$ since, ideally, the quantity calculated should be independent of chosen aperture size. However, it is quite possible that the spread in these values may in part be due to non-ideallized behavior, i.e. that the coma flux does not drop off as ${\rho}^{-1}$. Residuals from the removal of inertial background sources from the co-added image may contribute to the $Af\rho$ uncertainty as well. We measure $Af\rho$ as 432 $\pm$ 21 cm during the 2014 Jan.\ visit, 726 $\pm$ 40 cm during the Jul.\ visit, and 724 $\pm$ 40 cm during the Sep.\ visit. These $Af\rho$ values are provided for comparison with each other. The values are not corrected for phase angle effects since these effects are poorly constrained for infrared wavelengths, nor do the uncertainties include the component from the uncertainty in the absolute photometric calibration. We report these values in Table~\ref{table:obs} and show how $Af\rho$ varied with heliocentric distance in Figure~\ref{fig:activityvstime}. The $Af\rho$ value reported by \citet{CBET...3888..1} from observations in late May falls between our $Af\rho$ values when using the same phase-angle correction method\footnote{http://asteroid.lowell.edu/comet/dustphase.html}. Because the phase-angle correction at 3.4$\mu$m is not well-constrained for comet dust, we hesitate to make any further direct comparisons between the optical and infrared $Af\rho$ values when they are not simultaneous. 

%Conversion of flux to dust. afp take magnitude, solar reflectance. assume band 1 dominated by reflected light. 11 aperture.

%physical size of aperture varies with heliocentric/geocentric distance.

%uncertainty comes from dispersion in aperture values - supposed to be independent of aperture size but it's not.

%HWHM is 3, radius of 9 aperture is 3 times this.

\subsection{Anomalous Emission at 4.6~$\mu$m}

\begin{figure}[h]
\centering
\includegraphics[totalheight=10cm]{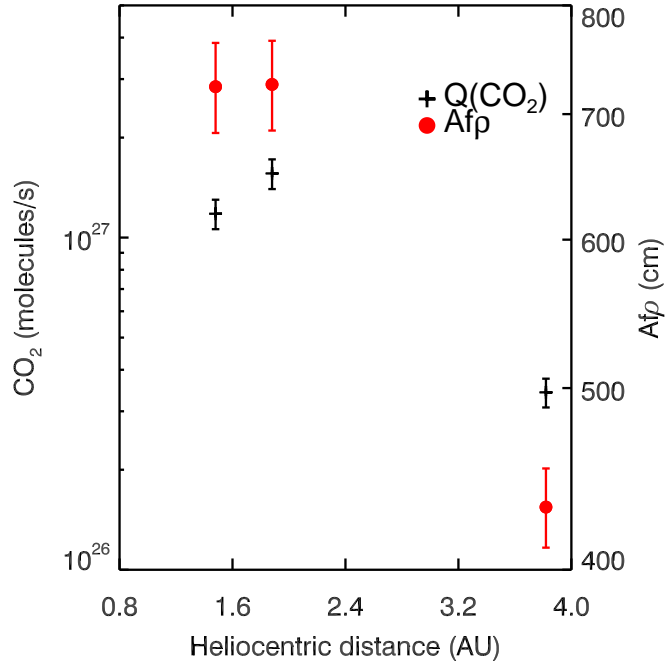}
\figcaption{The variation of Af$\rho$ and CO$_{2}$ production over time as observed by NEOWISE.}
\label{fig:activityvstime}
\end{figure}

At 4.6~$\mu$m the observed flux exceeds that predicted by the combined reflected and thermal light models.  In 2014 Jan.\ we find that the excess emission is at the 3~$\sigma$ level, while it drops to 2.5~$\sigma$ and 1.1~$\sigma$ in the Jul.\ and Sep.\ visits, respectively. We note that the uncertainties are dominated by systematics that correlate with the signal at 3.4~$\mu$m and thus the uncertainties at 4.6~$\mu$m are likely overestimated. The September band-excess detection is notably weak. However, the 3.4$\mu$m and 4.6$\mu$m relative photometry are better constrained, since the relative uncertainties are $\sim 7\%$ \citep{2012wise.rept....1C}, considerably less than the uncertainties listed in Table 1. The NEOWISE bandpass at 4.6~$\mu$m contains emission features from CO$_{2}$ ($\nu_{3}$ band) at 4.26~$\mu$m and CO (v=1-0 band) at 4.67~$\mu$m. Both species have sufficiently long photodissociation lifetimes to be responsible for the observed excess flux. We therefore  interpret the excess emission as being due to optically thin gaseous emission and convert the excess flux, $F$, to an average column density, $\langle N \rangle$, using equation~\ref{eq:gas}:

%Error bars include systematic effects that may correlate with the W1 signal and thus the uncertainties in the band 2 excess are likely overestimated.

\begin{equation}
\langle N \rangle = F 4 \pi \Delta^{2} \frac{\lambda}{h c} \frac{r_{h}^{2}}{g} \frac{1}{\pi \rho^{2}}
\label{eq:gas}
\end{equation}

where symbols have the same meanings as in equation~\ref{eq:afp}, $\lambda$ is the wavelength of observation, $h$ is Planck's constant, $c$ is the speed of light, and $g$ is the fluorescence efficiency for the chosen gas species.  $F$ is the excess flux density in the 4.6 $\mu$m bandpass is integrated over the CO/CO$_2$ band, after the in-band dust signal contribution has been removed, as described in \citet{2008AJ....136.1127P}. We are unable to distinguish between CO and CO$_{2}$ emission with just the NEOWISE data since the bandpass spans both features and therefore present two scenarios in Table~\ref{table:obs} where 100\% of the excess flux is due to CO emission or it is entirely due to  CO$_{2}$ emission. We assume fluorescence efficiencies at a heliocentric distance of 1~AU for the CO v=1-0  and the CO$_{2} ~ \nu_{3}$ bands of 2.46 $\times$ 10$^{-4}$ s$^{-1}$ and 2.86 $\times$ 10$^{-3}$ s$^{-1}$, respectively \citep{1983A&A...126..170C}. We then use the average column densities to calculate the production rates, $Q$ in molecules s$^{-1}$, using equation~\ref{eq:gasrates}:

\begin{equation}
Q = \langle N \rangle 2 \rho v \times 10^{5}
\label{eq:gasrates}
\end{equation}

in which $\langle N \rangle$ and $\rho$ have their previous definitions, $v$ is the ejection velocity of the gas in ~km~s$^{-1}$ and $10^{5}$ is a conversion factor \citep{2008AJ....136.1127P}. We take the ejection velocity to be 0.6~km~s$^{-1}$ during the 2014 Jan.\ visit when the comet was at a heliocentric distance of 3.82~AU and assume that the velocity scales as $\sqrt{r_{H}}$ \citep{1982come.coll...85D}. The production rates calculated are given in Table~\ref{table:obs} and show that gas production increased overall by a factor of $\sim$~3.5 between 2014 Jan.\ and Sep.. For these heliocentric distances, this kind of increase, proportional to $\sim r^{-1.8}$ is comparable to a Q$_{CO2} \sim r^{-2}$, possibly coinciding with a fixed source of constant area \citep{2004come.book..317M}. There was a decrease between the Jul.\ visit and the Sep.\ visit which may be attributable to a decrease in overall activity.  Figure~\ref{fig:activityvstime} shows the evolution of the $Af\rho$ quantity and CO$_{2}$ production as the comet moves toward perihelion. There are comparatively few published H$_2$O values for Siding Spring, but Q$_{H2O}$ was measured to be $\sim 2 \times 10^{27}$ molecules s$^{-1}$ when the comet was at $\sim 2.5$ AU \citep{CBET...3888..1}. This would suggest a lower bound of Q$_{CO2}$/Q$_{H2O}$  of $15\%$ assuming water production increased with decreasing heliocentric distance from 3.8 to 2.5 AU.

%Assumption of v.

%g factor. mol s.

%dust albedo? 0.08. twice what it is the optical. It's the A in Afp.

%Barovsky - REFERENCE? velocity. 0.6 km/s in January, 0.87 km/s in July, September: 1.0. 0.6 for the first value. delsemme 1982. Bobrovnikoff/Delsemme law.

%Page 1134 of Pittichova 2008. g factor. Assumed velocity for gas, convert column density into production rate in equation 5.

%For method reference: \citep{2008AJ....136.1127P}

% estimated flux at 4.6 microns, actual, sigma of detection
% Jan: 0.175765, real = 0.5670687 pm 0.13. Sigma = 3.01
% Jul: 9.25304, real = 20.984192 pm 4.62. Sigma = 2.54
% Sep: 55.9891, real = 74.158401 pm 16.3. Sigma = 1.1

\section{Conclusions}

We have used data from the NEOWISE reactivation mission to monitor the activity of comet Siding Spring over 9 months as it approaches Mars on a trajectory that will bring it within $\sim$135,000 km of the planet. Our conclusions from this work are as follows:

\begin{enumerate}

\item{NEOWISE observed comet Siding Spring to be active from January to September, 2014. The quantity $Af\rho$ and CO/CO$_{2}$ production rates initially increased as the comet reached a heliocentric distance of 1.88 AU but then decreased slightly even as the comet moved further inwards.}

\item{The activity of the comet decreased between July and September, possibly due to depletion of volatile deposits on or near the surface.}

\item{The decrease in activity suggests the risk to assets at Mars was reduced since activity diminished as the comet approached Mars.}

\end{enumerate}

\section{Acknowledgments}

This publication makes use of data products from NEOWISE, which is a project of the Jet Propulsion Laboratory/California Institute of Technology, funded by the National Aeronautics and Space Administration. This research has made use of the NASA/ IPAC Infrared Science Archive, which is operated by the Jet Propulsion Laboratory, California Institute of Technology, under contract with the National Aeronautics and Space Administration. RS gratefully acknowledges support from the NASA Postdoctoral Program.

\end{document}